\documentclass[aps,letterpaper,nofootinbib,preprint,showpacs,amsmath,amssymb,pdftex11pt]{revtex4-2}%
\usepackage{latexsym}%
\usepackage[dvips,dvipdfmx,hypertex]{hyperref}%

\def\cE{{\mathcal E}}

\def\cH{{\mathcal H}}

\DeclareMathAlphabet{\mathpzc}{OT1}{pzc}{m}{it}

\newcommand{\beq}{\begin{equation}}
\newcommand{\beqn}{\begin{equation}\nonumber}
\newcommand{\eeq}{\end{equation}}
\newcommand{\bea}{\begin{eqnarray}}
\newcommand{\bean}{\begin{eqnarray}\nonumber}
\newcommand{\eea}{\end{eqnarray}}

\begin{document}

\begin{center}
{\bf{\Large Proper time Quantization of a Thin Shell\footnote{This essay was selected for the 
First Award in the 2022 Gravity Research Foundation Essay Competition.}}}
\bigskip
\bigskip

Cenalo Vaz
\bigskip

{\it University of Cincinnati Blue Ash,}\\
{\it Cincinnati, Ohio 45236, USA}\\
{email: Cenalo.Vaz@UC.Edu}
\medskip

\end{center}
\bigskip
\medskip

``Time'' has different meanings in classical general relativity and in quantum theory. While all choices of a time function 
yield the same local classical geometries, quantum theories built on different time functions are not unitarily equivalent. 
This incompatibility is most vivid in model systems for which exact quantum descriptions in different time variables are available.
One such system is a spherically symmetric, thin dust shell. In this essay we will compare the quantum theories of the shell built 
on proper time and on a particular coordinate time. We find wholly incompatible descriptions: whereas the shell quantum mechanics 
in coordinate time admits {\it no} solutions when the mass is greater than the Planck mass, its proper time quantum mechanics {\it only} 
admits solutions when the mass is greater than the Planck mass. The latter is in better agreement with what is expected from 
observation. We argue that proper time quantization provides a superior approach to the problem of time in canonical quantization.

\vfill\eject\

Canonical quantization requires a foliation of a globally hyperbolic spacetime by a sequence of spatial hypersurfaces. 
One can foliate the spacetime in an infinite number of ways, each giving rise to a unique ordering of the hypersurfaces and hence to 
a unique time function. Classical general relativity is independent of the foliation in that all of the foliations describe the same 
local geometries but the situation is different in the quantum theory, which relies on time as an observer independent, 
classical variable (Newton's absolute time). Quantum theories built on different time variables are not unitarily equivalent. In relativity,
proper time assumes the role of Newton's absolute time in the sense that it satisfies the condition of being observer independent. It 
is therefore worth comparing the quantum descriptions of model systems in proper time and in coordinate time where exact solutions are 
possible. 

One such system, that is also of considerable physical interest, is a self-gravitating dust shell of infinitesimal thickness. On the 
classical level, the shell has just one degree of freedom and is completely described by its radius, $R(t)$ and 
its conjugate momentum, $P(t)$. Yet, various versions of it form a rich enough collection of physical systems to describe the final 
stages of gravitational collapse, Hawking radiation and the formation (or avoidance) of gravitational singularities \cite{visser04,
xu06,vachaspati08,vachaspati09,paranjape09,bardeen14,ziprick16,binetruy18,baccetti20}. There are three distinct time variables present 
in the problem, each of which is ``natural'' in some setting. These are (i) the time coordinate appropriate to the interior of the 
shell, (ii) the time coordinate in the exterior of the shell and (iii) the comoving (proper) time of the shell. The shell is described 
by one conservation law that may be construed as a first integral of an equation of motion. This is obtained by applying the junction 
conditions of Israel-Darmois-Lanczos \cite{israel66,darmois27,lanczos24}. If the shell is collapsing in a vacuum, {\it i.e.,} the 
interior is taken to be Minkowski spacetime and the exterior a Schwarzschild spacetime of ADM mass $M$, one finds
\beq
M = m\sqrt{1+R_\tau^2} - \frac{Gm^2}{2R},
\label{vacuumconst}
\eeq
where $m$ is a global constant representing the proper mass of the shell, $R(\tau)$ is the shell radius, $\tau$ is the shell proper 
time and the subscript indicates a derivative with respect to that variable \cite{poisson19}. The Minkowski time ($T$) and the 
Schwarzschild time ($t$) are related to the proper time by
\beq
\frac{dT}{d\tau} = \sqrt{1 + R_\tau^2},~~ \frac{dt}{d\tau} = \frac{\sqrt{B + R_\tau^2}}B
\label{timeders2}
\eeq
where $B = 1-2GM/R$. The ADM mass, $M$, clearly represents an energy, but the constraint is expressed in terms of the velocities and 
there is some ambiguity in constructing the Hamiltonian because one does not know {\it \`a priori} in which of the three time variables 
$M$ evolves the system. For example, if $M$ is taken to evolve the system in proper time, one arrives at \cite{berezin88}
\beq
H = m \cosh \frac pm - \frac{Gm^2}{2R}
\eeq
(The corresponding operator has derivatives of all orders but was shown to possess a positive self adjoint extension in \cite{hajicek92b}.)
On the other hand,  if the ADM mass is taken to evolve the system in the time coordinate of the interior of the shell it leads to the 
Hamiltonian
\beq
H = - p_T = \sqrt{p^2 + m^2} - \frac{Gm^2}{2R},
\label{haminT}
\eeq
This ambiguity has nothing to do with quantum mechanics and arises because the shell equation was constructed from junction conditions and 
not an action principle. However, attempts at recovering \eqref{vacuumconst} from a more fundamental action principle have not been 
successful \cite{louko98,hajicek01}. In what follows we will choose \eqref{haminT} to be the canonical choice (as did the authors of 
\cite{hajicek92a}). From $H$, we can construct an effective action for the shell \cite{vachaspati08,vachaspati09}
\beq
S = \int dT \left[-m \sqrt{1-R_T^2} + \frac{Gm^2}{2R}\right]
\eeq
and transform this action to proper time with the help of \eqref{timeders2},
\beq
S = \int d\tau \left[-m + \frac{Gm^2}{2R} \sqrt{1+R_\tau^2}\right].
\label{actiontau}
\eeq
In this way, we arrive at the Hamiltonian for the evolution of the shell in proper time,
\beq
\cH = - P_\tau = m - \sqrt{f^2-P^2},
\label{hamintau}
\eeq
where $f(R) = G m^2/2R$. The Hamiltonian $\cH$ is remarkably similar in structure to the superhamiltonian obtained in \cite{vaz01} for a marginally bound 
dust ball in a midisuperspace quantization of the Einstein-Dust system \cite{ltb}. 

We now have two Hamiltonians describing the same system, the 
first ($H$) evolving the shell in (the interior) coordinate time and the second ($\cH$) evolving the shell in comoving time, so we 
can examine and compare the quantum descriptions of the shell in these two times. We will base the quantum theories on their 
corresponding superhamiltonians, which are quadratic in the momenta and from which the equations of motion can be derived in each case. 
For $p_T$ in \eqref{haminT} we have
\beq
h_T = (p_T - f)^2 - p^2 -m^2 = 0
\label{superhT}
\eeq
and for $P_\tau$ in \eqref{hamintau},
\beq
h_\tau = (P_\tau + m)^2 + P^2 - f^2 = 0,
\label{superh1}
\eeq
These give the wave equations 
\beq
\left[\left(-i\partial_T - f\right)^2 +\partial_R^2 - m^2\right]\Psi(T,R) = 0
\eeq
and 
\beq
\left[(-i \partial_\tau + m)^2 - \partial_R^2 - f^2 \right]\Psi(\tau,R) = 0
\eeq
respectively. It is straightforward that they respectively yield the dynamical equations \eqref{haminT} and \eqref{hamintau} 
in the classical limit. Notice that the wave equation is hyperbolic in (interior) coordinate time 
and elliptic in proper time.

The quantum theory of \eqref{superhT} was discussed in detail in \cite{hajicek92a}, so we will first consider the quantum theory of the 
shell in proper time given in \eqref{superh1}. For any two solutions of the wave equation, $\Phi$ and $\Psi$, there is a conserved bilinear 
current density,
\beq
J_i = - \frac i2 \Phi^* \overleftrightarrow{\partial_i}\Psi + m \delta_{i\tau}\Phi^*\Psi,~~ i \in \{\tau,R\},
\label{conscurtau}
\eeq
the time component of which specifies a physical inner product
\beq
\langle \Phi,\Psi\rangle = \int_0^\infty dR\left[- \frac i2 \Phi^* \overleftrightarrow{\partial_\tau}\Psi + m \Phi^*\Psi\right],
\label{inprod}
\eeq
sometimes referred to as the ``charge'' form in analogy with the classical charged field. The charge form is positive semi-definite 
as long as $\cE<m$ because the equation is elliptic and it may be taken to represent a probability density. This is a generic feature of 
the proper time quantization. Therefore, with \eqref{inprod} we obtain an inner product space that can be extended to a separable Hilbert 
space by Cauchy completion \cite{reed80,wald94}. We confine our attention to stationary states,
\beq
\Psi(\tau,R) = e^{-i\cE \tau}\psi(R),
\label{statstate}
\eeq
which leads to the following radial equation:
\beq
\psi''(R) - \left[(m-\cE)^2 - \frac{\mu^4}{4R^2} \right]\psi(R) = 0,
\label{radialeqn}
\eeq
where $\mu$ is the ratio of the shell mass to the Planck mass, $\mu = m/m_p$. The general solution of the radial equation in 
\eqref{radialeqn} behaves as $e^{\pm (m-\cE)R}$ at large $R$, and can be expressed as a linear combination of Bessel functions 
of the first and second kind,
\beq
\psi(R) = \sqrt{R}\left[C_1 J_\sigma(-i\alpha R) + C_2 Y_\sigma(-i\alpha R)\right],
\eeq
where we let $\alpha = m-\cE > 0$ and $\sigma = \frac 12 \sqrt{1-\mu^4}$. Normalizability, according to \eqref{inprod}, 
requires $\Psi$ to fall off exponentially at infinity, with implies that $C_1 = i C_2$.  Thus $\phi(R)$ is Hankel's Bessel function 
of the third kind and the exact solution is
\beq
\Psi(\tau,R) = C e^{-i\cE\tau}\sqrt{R} H^{(2)}_\sigma(-i\alpha R),
\label{exactsoltau}
\eeq
where $C$ is an overall constant. As $R\rightarrow 0$, $\Psi(\tau,R)$ behaves as
\beq
\Psi(\tau,R) \sim \left\{\begin{matrix}
C\sqrt{R}e^{-i\cE\tau}\left[\frac i\pi \Gamma(\sigma)\left(\frac{\alpha R}2\right)^{-\sigma} e^{\frac{i\pi\sigma}2}
+ \frac{(1-i\cot\pi\sigma)}{\Gamma(1+\sigma)} \left(\frac{\alpha R}2\right)^{\sigma}e^{-\frac{i\pi\sigma}2}\right],~~ \sigma\neq 0\cr\cr
\frac{2C}\pi \sqrt{R} \left[\gamma +\ln \left(\frac{\alpha R}2\right)\right],~~ \sigma = 0
\end{matrix}\right.
\label{nearzero}
\eeq
where $\gamma$ is Euler's constant. The behavior of these solutions near the center will depend on the mass ratio, $m/m_p=\mu$. If the 
shell mass is less than the Planck mass, $\mu < 1$, then $0\leq\sigma<1/2$ is real (we exclude the case $m=0$ because our construction 
is valid only for a timelike shell), but if the shell's rest mass is greater than the Planck mass, $\sigma$ is imaginary. 

Consider two stationary solutions, $\Phi_{\cE'}$ and $\Psi_\cE$, with energies $\cE'$ and $\cE$ respectively. 
The inner product \eqref{inprod} becomes 
\beq
\langle\Phi_{\cE'},\Psi_\cE\rangle = \frac 12\left[2m-(\cE+\cE')\right]e^{-i(\cE-\cE')\tau}\int_0^\infty dR~ \phi^*_{\cE'}\psi_\cE
\eeq
and by the equation of motion, we have 
\bea
\phi^*_{E'}\psi_\cE'' - \left((m-\cE)^2 - f^2\right)\phi^*_{\cE'}\psi_\cE &=& 0\cr\cr
\psi_{E}\phi^{*''}_{\cE'} - \left((m-\cE')^2 - f^2\right)\psi_\cE\phi^*_{\cE'} &=& 0
\eea
Subtracting the second from the first,
\beq
\phi^*_{\cE'}\psi_\cE'' - \psi_{\cE}\phi^{*''}_{\cE'} = (\phi^*_{\cE'}\overleftrightarrow{\partial_R}\psi_\cE)' = 
(\cE-\cE')(\cE+\cE'-2m)\phi^*_{\cE'}\psi_\cE 
\eeq
and it follows that the inner product is a boundary term,
\beq
\langle\Phi_{\cE'},\Psi_\cE\rangle =  \left.\frac{i J_R}{(\cE'-\cE)}\right|_0^\infty
\label{inprod2}
\eeq
where
\beq
J_R = -\frac i2 e^{-i(\cE-\cE')\tau}\phi^*_{\cE'}\overleftrightarrow{\partial_R}\psi_\cE
\eeq
is the radial component of the $U(1)$ current in \eqref{conscurtau}. The exponential fall off of our wave function at infinity 
ensures that $J_R$ vanishes there. The inner product therefore depends only on the value of the radial current at the origin. 

To guarantee orthonormality of the wave functions, we must require that the inner product of two wave functions of different 
energies vanishes. In particular, this means that $J_R$ should vanish at the origin when $\cE\neq \cE'$. Evaluating $J_R$, using 
the behavior of the solutions in \eqref{nearzero}, we find 
\beq
J_R \sim \left\{\begin{matrix}
\frac{|C|^2}{\sin\pi\sigma} \left[\left(\frac{m-\cE}{m-\cE'}\right)^\sigma-\left(\frac{m-\cE'}{m-\cE}\right)^\sigma\right],~~ 
\sigma\neq 0\cr\cr
\frac{2|C|^2}{\pi^2}\left[4in\pi + 2 \ln \left(\frac{m-\cE'}{m-\cE}\right)\right],~~ \sigma=0
\end{matrix}\right.
\eeq
If $\sigma$ is real ($m\leq m_p$) $J_R$ does not vanish, therefore there is no orthogonal set of solutions in this case. However, if 
the mass of the shell is greater than the Planck mass then $\sigma$ is imaginary and letting $\sigma = i\beta$,
\beq
J_R \sim \frac{|C|^2}{\sinh\pi\beta} \left[\left(\frac{m-\cE}{m-\cE'}\right)^{i\beta}-\left(\frac{m-\cE}{m-\cE'}\right)^{-i\beta}\right]
\eeq
vanishes if 
\beq
\frac{m-\cE}{m-\cE'} = e^{n\pi/\beta}
\label{ratio}
\eeq
for any integer $n$. Now the energy operator commutes with the superhamiltonian and there is proof of the positivity of energy
in General Relativity, so it is reasonable to exclude negative energy states and take the ground state to have zero energy.
Then \eqref{ratio} amounts to the energy spectrum,
\beq
\cE_n = m \left(1-e^{-n\pi/\beta}\right),
\label{spectrumtau}
\eeq
where $n$ is a positive integer.

Thus, from the comoving observer's point of view, there is a quantum theory of the shell but only for masses larger than the Planck mass.
Both the wave function and the $U(1)$ charge current density vanish at the center and are well behaved everywhere. 
The energy spectrum is discrete and, near the center, each energy eigenfunction is a combination of an infalling wave and an outgoing wave,
\beq
\Psi_n(\tau,R) \sim \frac{ie^{-\frac{\pi\beta}2}}\pi\Gamma(i\beta) C\sqrt{R}\left[\underbrace{e^{-i(\cE_n\tau + \beta \ln 
\frac{\alpha_n R}2)}}_{\text{infalling}} + \frac{\pi}{\beta\Gamma^2(i\beta)\sinh\pi\beta} \underbrace{e^{-i(\cE_n\tau-\beta \ln 
\frac{\alpha_n R}2)}}_{\text{outgoing}}\right]
\label{nearzerowf}
\eeq
with only a relative phase shift that depends on $m/m_p$.

Next, consider the shell quantum mechanics in (internal) coordinate time. An in-depth analysis of this system was given in \cite{hajicek92a}, 
but the wave function was subjected to additional conditions because of the analogy with the classical, spherically symmetric, charged 
Klein-Gordon field in an external Coulomb potential, to which \eqref{superhT} is identical. The energy and momentum forms were required to 
also vanish at the origin. However, positivity of the energy and a well behaved $U(1)$ current are sufficient to build a separable Hilbert 
space and the additional conditions select a subset of the otherwise well defined Hilbert space, placing completeness in doubt. To compare 
the result with the proper time quantum theory, we will drop the additional conditions in the following.

The radial equation for positive energy stationary states reads,
\beq
\psi'' + \left[(E^2-m^2) + \frac{\mu^2 E}R + \frac{\mu^4}{4R^2}\right]\psi = 0,
\eeq
and one can show, as before, that the charge form bears the same relationship to the radial charge current as \eqref{inprod2}. This time, 
however, the radial charge current does not vanish for two states with different energies at $R=0$ when $\mu>1$. Therefore there are no 
solutions when $m>m_p$. When $\mu < 1$, the radial charge current can be made to vanish and orthogonal states can be defined. Scattering 
states ($E>m$) are given by the Kummer function as indicated in \cite{hajicek92a}. Bound states ($E<m$) can be given in terms of the confluent 
hypergeometric function. We obtain
\beq
\Psi^\pm_n(\tau,R) = C R^{\frac 12\pm\sigma} e^{-\alpha^\pm_n R}U(-n, 1\pm 2\sigma,2\alpha^\pm_n~R)
\label{exactsolT}
\eeq
where $U(a,b,x)$ is the confluent hypergeometric function, $n$ is a whole number, $\alpha^\pm_n = \sqrt{m^2-E_n^{\pm 2}}$, $\sigma=\frac 12 
\sqrt{1-\mu^4}$ and $E^\pm_n$ is given by
\beq
E^\pm_n = \frac{2 m \left(\lambda_\pm + n\right)}{\sqrt{\mu^4 + 4(\lambda_\pm +n)^2}}
\label{spectrumT}
\eeq
where $\lambda_\pm =\frac 12\left(1\pm \sigma\right)$. The subset $\{\psi^-_n\}$ is eliminated if the classical field energy-momentum is also 
required to vanish at the center, but then completeness of the subset $\{\psi^+_n\}$ must be explicitly verified.

We find a stark difference in the quantum descriptions of the shell in coordinate time and proper time, in fact the quantum theories are 
incompatible. In particular, the quantum theory in Minkowski time has no solutions for masses greater than the Planck mass, which appears 
to be in direct opposition to reality. It is satisfying therefore that the proper time quantization addresses just this physically relevant 
mass regime. 

In general, proper time quantization seems to enjoy several advantages over coordinate time quantizations, the most important being that 
it satisfies the basic requirement of observer independence. It is therefore ``democratic'' in regard to coordinate time (physical observers) 
in that all coordinate time variables would be functions of the phase space (in the simple case of the shell these are given by 
\eqref{timeders2}) as are the spatial coordinates. Thus they would all be operator valued and we would be able to speak of time intervals 
only in terms of averages. In the proper time formulation, these averages can be calculated and fluctuations about them quantified because 
the Schroedinger equation yields a conserved, positive semi-definite inner product. (The same would be true of the metric components, implying 
that one must always deal with fuzzy local geometries.) A proposal for generically introducing a privileged frame and proper time foliation 
was made by Brown and Kucha\v r by the introduction of tenuous, incoherent dust \cite{kuchar94}.

\end{document}